# Parameterized Low-distortion Embeddings - Graph metrics into lines and trees

Michael Fellows [*]   Fedor V. Fomin[†]   Daniel Lokshtanov[†]   Elena Losievskaja [‡]

Frances Rosamond [*]   Saket Saurabh[†]

## Abstract

We revisit the issue of low-distortion embedding of metric spaces into the line, and more generally, into the shortest path metric of trees, from the parameterized complexity perspective. Let $M = M(G)$ be the shortest path metric of an edge weighted graph $G = (V, E)$ on $n$ vertices. We describe algorithms for the problem of finding a low distortion non-contracting embedding of $M$ into line and tree metrics.

- We give an $O(nd^4(2d+1)^{2d})$ time algorithm that for an *unweighted* graph metric $M$ and integer $d$ either constructs an embedding of $M$ into the line with distortion at most $d$, or concludes that no such embedding exists. We find the result surprising, because the considered problem bears a strong resemblance to the notoriously hard BANDWIDTH MINIMIZATION problem which does not admit any FPT algorithm unless an unlikely collapse of parameterized complexity classes occurs. The running time of our algorithm is a significant improvement over the best previous algorithm of Bădoiu *et al.* [SODA 2005] that has a running time of $O(n^{4d+2}d^{O(1)})$.

- We show that our algorithm can also be applied to construct small distortion embeddings of *weighted* graph metrics. The running time of our algorithm is $O(n(dW)^4(2d+1)^{2dW})$ where $W$ is the largest edge weight of the input graph. To complement this result, we show that the exponential dependence on the maximum edge weight is unavoidable. In particular, we show that deciding whether a weighted graph metric $M(G)$ with maximum weight $W < |V(G)|$ can be embedded into the line with distortion at most $d$ is NP-Complete for every fixed rational $d \geq 2$. This rules out any possibility of an algorithm with running time $O((nW)^{h(d)})$ where $h$ is a function of $d$ alone.

- We generalize the result on embedding into the line by proving that for any tree $T$ with maximum degree $\Delta$, embedding of $M$ into a shortest path metric of $T$ is FPT, parameterized by $(\Delta, d)$. This result can also be viewed as a generalization (albeit with a worse running time) of the previous FPT algorithm due to Kenyon, Rabani and Sinclair [STOC 2004] that was limited to the situation where $|G| = |T|$.

---

[*]University of Newcastle, Newcastle, Australia. {michael.fellows,frances.rosamond}@newcastle.edu.au
[†]Department of Informatics, University of Bergen, Bergen, Norway. {fedor.fomin,daniello,saket}@ii.uib.no
[‡]Dept. of Computer Science, University of Iceland, Iceland. elenal@hi.is

# 1  Introduction

Given an undirected graph $G = (V, E)$ together with a weight function $w$ that assigns a positive weight $w(uv)$ to every edge $uv \in E$, a natural metric associated with $G$ is $M(G) = (V, D_G)$ where the distance function $D_G$ [1] is the weighted shortest path distance between $u$ and $v$ for each pair of vertices $u, v \in V$. We call $M(G)$ as the (weighted) *graph metric* of $G$. If $w(uv) = 1$ for every edge $uv \in E$, we say that $M(G) = (V, D_G)$ is an *unweighted graph metric*. For a subset $S$ of $V(G)$, we say that $M[S] = (S, D'')$ (where $D''$ is $D$ restricted to $S^2$) is the submetric of $M(G)$ induced by $S$. Given a graph metric $M$ and another metric space $M'$ with distance functions $D$ and $D'$, a mapping $f : M \to M'$ is called an *embedding* of $M$ into $M'$. The mapping $f$ has *contraction* $c_f$ and *expansion* $e_f$ if for every pair of points $p, q$ in $M$, $D(p, q) \leq D'(f(p), f(q)) \cdot c_f$ and $D(p, q) \cdot e_f \geq D'(f(p), f(q))$ respectively. We say that $f$ is *non-contracting* if $c_f$ is at most 1. A non-contracting mapping $f$ has *distortion* $d$ if $e_f$ is at most $d$.

Embedding a graph metric into a simple metric space like the real line has proved to be a useful tool in designing algorithms in various fields. A long list of applications given in [8] includes approximation algorithms for graph and network problems, such as sparsest cut, minimum bandwidth, low-diameter decomposition and optimal group steiner trees, and online algorithms for metrical task systems and file migration problems. These applications often require algorithms for finding low distortion embeddings, and the study of the algorithmic issues of metric embeddings has recently begun to develop [1, 2, 3, 11]. For example, Bǎdoiu *et al.* [1, 3] describe approximation algorithms and hardness results for embedding general metrics into the line and tree metrics respectively. In particular they show that the minimum distortion for a line embedding is hard to approximate up to a factor polynomial in $n$ even for weighted trees with polynomial spread (the ratio of maximum/minimum weights). Hall and Papadimitriou [9] studied the hardness of approximation for bijective embeddings. Independently from the algorithmic viewpoint, the problem of finding a low-distortion embedding between metric spaces is a fundamental mathematical problem [10, 12] that has been studied intensively.

In many applications one needs the distortion of the required embedding to be relatively small. Hence it is natural to study the algorithmic issues related to small distortion embeddings within the framework of parameterized complexity [6, 7, 13]. This paradigm associates a natural secondary measurement to the problem and studies the algorithmic behavior of the problem in terms of the associated measurement, called the *parameter*. In this paper we consider a natural parameter, the *distortion* $d$, and consider the feasibility of having an algorithm of time complexity $g(d) \cdot n^{O(1)}$ for the problem of embedding weighted graph metrics into the line with distortion at most $d$.

What would one expect about the complexity of embedding an *unweighted* graph metric into the line, parameterized by the distortion $d$? At a glance, the problem seems to closely resemble the BANDWIDTH MINIMIZATION problem. In the BANDWIDTH MINIMIZATION problem one is given a graph $G = (V, E)$ and asked to find a bijective mapping $f : V \to \{1, \ldots, n\}$, for which the bandwidth, i.e. $b = \max_{(u,v) \in E} |f(u) - f(v)|$, is minimized. This problem is known to be $W[t]$-hard for all $t \geq 1$ [4, 5], when parameterized by $b$. Unless an unlikely collapse of parameterized complexity classes occurs, this rules out any possibility of having an algorithm with running time $g(b) \cdot n^{O(1)}$ for BANDWIDTH MINIMIZATION and thus the algorithm of Saxe [14] running in time $O(4^b n^{b+1})$ is essentially the best possible. Previous to this paper, the best algorithm (by Bǎdoiu *et al.* [2]) to decide whether an unweighted graph metric can be embedded into the line with distortion at most $d$ has a running time where $d$ appears in the exponent of $n$, that is $O(n^{4d+2} \cdot d^{O(1)})$. Bǎdoiu *et al.* [2] mention that for exact algorithms, this problem seems to have many similarities with the BANDWIDTH MINIMIZATION problem. They remark that "... our exact algorithm for computing the distortion is based on the analogous result for the bandwidth problem by Saxe [14].". Because of the apparent similarity to the notoriously hard bandwidth problem, it is very surprising that, in fact, this fundamental problem of embedding unweighted graph metrics into the line turns out to be fixed parameter tractable (FPT).

---
[1] We also denote the distance function $D_G$ by $D$ if the graph in consideration is clear from the context.

**Theorem 1.1.** *Given an unweighted graph $G = (V, E)$ we can decide whether $M(G)$ can be embedded into the real line with distortion at most $d$ in time $O(nd^4(2d+1)^{2d})$.*

The running time of the algorithm is linear for every fixed $d$ and clearly improves the running time of the previously known algorithm. In fact, one can apply Theorem 1.1 in order to check whether the unweighted graph metric can be embedded into the line with distortion at most $\lg n / \lg \lg n$ in time polynomial in $n$.

Having coped with the unweighted case, we return to the study of low distortion embeddings of weighted graph metrics into the line. We show that if the maximum weight of any edge is bounded by $W$, then we can modify the algorithm presented in Theorem 1.1 to give an algorithm to decide whether $M(G)$ can be embedded into the line with distortion at most $d$ in time $O(n(dW)^4(2d+1)^{2dW})$. However the weights in a graph metric do not need to be small, and hence this algorithm is not sufficient to give a $g(d) \cdot n^{O(1)}$ time algorithm for the problem of embedding weighted graph metrics into the line. Can such an algorithm exist? Unfortunately, it turns out that our $O(n(dW)^4(2d+1)^{2dW})$ algorithm essentially is the best one can hope for. In fact, our next result rules out not only any possibility of having an algorithm with running time of the form $g(d) \cdot n^{O(1)}$, but also any algorithm with running time $(nW)^{h(d)}$, where $h$ only depends on $d$.

**Theorem 1.2.** *Deciding whether a weighted graph metric $M(G)$ with maximum weight $W < |V(G)|$ can be embedded into the line with distortion at most $d$ is NP-Complete for every fixed rational $d \geq 2$.*

Another direction for generalizing Theorem 1.1 is to look for other simple topologies or host metrics for which an analogous result to Theorem 1.1 holds. Kenyon *et al.* [11] provided FPT algorithms for the *bijective* embedding of unweighted graph metrics into the metric of a tree with bounded maximum degree $\Delta$. The running time of their algorithm is $n^2 \cdot 2^{\Delta^{\alpha^3}}$ where $\alpha$ is the maximum of $c_f$ and $e_f$. An important point, observed in [2], is that constraining the embedding to be bijective (not just injective, as in our case) is crucial for the correctness of the algorithms from [11]. We complement the FPT result of Kenyon, Rabani and Sinclair [11] by extending our results to give an algorithm for the problem of embedding unweighted graph metrics into a metric generated by a tree with maximum degree bounded by $\Delta$, parameterized by distortion $d$ and $\Delta$.

**Theorem 1.3.** *Given a graph $G$, a tree $T$ with maximum degree $\Delta$ and an integer $d$ we can decide whether $G$ can be embedded into $T$ with distortion at most $d$ in time $n^2 \cdot |V(T)| \cdot 2^{O((5d)^{\Delta^{d+1}} \cdot d)}$.*

Why stop at bounded degree trees? Can our results be extended to yield FPT algorithms for low distortion embeddings into other, more complicated topologies? At a first glance, this seems to be the case. However, even a simple change in the topology of the host metric can change the behavior of the problem, or even make the problem completely intractable. For embedding into the line it is enough to control the local properties of the embedding whereas this is not sufficient for embedding into bounded degree trees. The techniques we use to cope with these difficulties do not look to be extendable to the problem of finding low distortion embeddings into cycles, an interesting open problem that seems to exhibit even more non-locality than that of embedding into bounded degree trees. A more dramatic example is that of low distortion embeddings of unweighted graph metrics into wheels (cycle with one additional vertex adjacent to all the vertices of a cycle). In fact, it turns out that deciding whether one can embed an unweighted graph metric into a wheel with distortion at most 2 is NP-complete, by a simple reduction from HAMILTONIAN CYCLE problem. Thus, the problems of embedding an unweighted graph metric into cycles and trees of unbounded degree remain interesting open problems.

## 2 Algorithms for Embedding Graph Metrics into the Line

### 2.1 Unweighted Graph Metrics into the Line

In this section we give an algorithm for embedding unweighted graph metrics into the line. We slightly abuse the terminology here by saying *embedding of a graph G* instead of embedding of the unweighted graph metric $M(G)$ of $G$. Before we proceed to the details of the algorithm we need a few observations that allow us to only consider a specific kind of embeddings. For a non-contracting embedding $f$ of a graph $G$ into the line, we say that vertex $u$ *pushes* vertex $v$ if $D(u,v) = |f(u) - f(v)|$.

**Observation 2.1.** [⋆] [2] *If $f(u) < f(v) < f(w)$ and $u$ pushes $w$, then $u$ pushes $v$ and $v$ pushes $w$.*

For an embedding $f$, let $v_1, v_2, \ldots, v_n$ be an ordering of the vertices such that $f(v_1) < f(v_2) < \ldots < f(v_n)$. We say that $f$ is *pushing* if $v_i$ pushes $v_{i+1}$, for each $1 \leq i \leq n-1$.

**Observation 2.2.** [⋆] *If $G$ can be embedded into the line with distortion $d$, then there is a pushing embedding of $G$ into the line with distortion $d$. Furthermore, every pushing embedding of $G$ into the line is non-contracting.*

**Observation 2.3.** *Let $f$ be a pushing embedding of a connected graph $G$ into the line with distortion at most $d$. Then $D(v_{i-1}, v_i) \leq d$ for every $1 \leq i \leq n$.*

By Observation 2.2, it is sufficient to work only with pushing embeddings. Our algorithm is based on dynamic programming over small intervals of the line. The intuition behind the algorithm is as follows. Let us consider a distortion $d$ embedding of $G$ into the line and an interval of length $2d+1$ of the line. First, observe that no edge can have one end-point to the left of this interval and one end-point to the right. This means that if there is a vertex $u$ embedded to the left of this interval and another vertex $v$ that has been embedded to the right, then the set of vertices embedded into the interval form an $u,v$-separator. Moreover, for each edge, its end-points can be mapped at most $d$ apart, and hence there is no edge with one end-point to the left of this interval and the other end-point in the rightmost part of this interval. Thus just by looking at the vertices mapped into an interval of length $2d + 1$, we deduce which of the remaining vertices of $G$ were mapped to the left and which were mapped to the right of this interval. This is a natural division of the problem into independent subproblems and the solutions to these subproblems can be used to find an embedding of $G$. Next we formalize this intuition by defining *partial embeddings* and showing how they are glued onto each other to form a distortion $d$ embedding of the input graph.

It is well known (and it follows from Observation 2.2) that there always exists an optimal embedding with all the vertices embedded into *integer coordinates* of the line. Without loss of generality, in the rest of this section we only consider pushing embeddings of this type. We also assume that our input graph $G$ is *connected*.

**Definition 2.4.** *For a graph $G$ and a subset $S \subseteq V(G)$, a partial embedding of $S$ is a function $f : S \to \{-(d+1), \ldots, d+1\}$. We define $S_f^{[a,b]}$, $-(d+1) \leq a \leq b \leq d+1$, to be the set of all vertices of $S$ which are mapped into $\{a, \ldots, b\}$ by $f$ (let us remark that this can be $\emptyset$). We also define $S_f^L = S_f^{[-(d+1), -1]}$ and $S_f^R = S_f^{[1, d+1]}$. For an integer $x$, $-(d+1) \leq x \leq d+1$, we put $S_f^x = S_f^{[x,x]}$. Finally, we put $L(f)$ ($R(f)$) to denote the union of the vertex sets of all connected components of $G \setminus S$ that have neighbors in $S_f^L$ ($S_f^R$).*

---

[2]Proofs of results labelled with [⋆] have been moved to the appendix due to space restrictions.

**Definition 2.5.** *A partial embedding $f$ of a subset $S \subseteq V(G)$ is called* feasible *if (1) $f$ is a non-contracting distortion $d$ embedding of $S$; (2) $L(f) \cap R(f) = \emptyset$; (3) Every neighbor of $S_f^0$ is in $S$; (4) if $R(f) = \emptyset$, then $S_f^{d+1}$ is nonempty; (5) if $L(f) = \emptyset$, then $S_f^{-(d+1)}$ is nonempty; (6) if $f(u) + 1 < f(v)$ and $S_f^{[f(u)+1, f(v)-1]} = \emptyset$, then $f(v) - f(u) = D(u,v)$. (Basically, $u$ pushes $v$.)*

The properties 1, 2, and 3 of this definition will be used to show that every distortion $d$ embedding of $G$ into the line can be described as a sequence of feasible partial embeddings that have been glued onto each other. Properties 4, 5 and 6 are helpful to bound the number of feasible partial embeddings.

**Definition 2.6.** *Let $f$ and $g$ be feasible partial embeddings of a graph $G$, with domains $S_f$ and $S_g$, respectively. We say that $g$ succeeds $f$ if (1) $S_f^{[-d,d+1]} = S_g^{[-(d+1),d]} = S_f \cap S_g$; (2) for every $u \in S_f \cap S_g$, $f(u) = g(u) + 1$; (3) $S_g^{d+1} \subseteq R(f)$; (4) $S_f^{-(d+1)} \subseteq L(g)$.*

The properties 1 and 2 describe how one can glue a partial embedding $g$ that has been shifted one to the right onto another partial embedding $f$. Properties 3 and 4 are employed to enforce "intuitive" behavior of the sets $L(f)$, $R(f)$, $L(g)$ and $R(g)$. That is, since $g$ is glued on the right side of $f$, everything to the right of $g$ should appear in the right side of $f$. Similarly, everything to the left of $f$ should be to the left of $g$.

**Lemma 2.7.** [⋆] *For every pair of feasible partial embeddings $f$ and $g$ of subsets $S_f$ and $S_g$ of $V(G)$ such that $g$ succeeds $f$, we have $R(f) = R(g) \cup S_g^{d+1}$ and $L(g) = L(f) \cup S_f^{-(d+1)}$.*

**Theorem 2.8.** *For every integer $d$, a graph $G$ has an embedding of distortion at most $d$ if and only if there exists a sequence of feasible partial embeddings $f_0, f_1, f_2, \ldots, f_t$ such that for each $0 \le i \le t-1$, $f_{i+1}$ succeeds $f_i$, and $L(f_0) = R(f_t) = \emptyset$.*

*Proof.* Let $f$ be a pushing embedding of $G$ with distortion $d$ which maps all vertices to integers greater than or equal to $-(d+1)$ and maps one vertex to $-(d+1)$. Let $t$ be the smallest integer such that $f(v) \le t + d + 1$ for every $v \in V$. For every $0 \le i \le t$, let $S_i$ be the set of vertices that $f$ maps to $\{i - (d+1), \ldots, i + d + 1\}$. We define $f_i : S_i \to \{-(d+1), \ldots, d+1\}$ to be $f_i(v) = f(v) - i$, $v \in S_i$. Then for every $i \le t-1$, $f_i$ is a feasible partial embedding, $f_{i+1}$ succeeds $f_i$, and $L(f_0) = R(f_t) = \emptyset$.

In the other direction, let $f_0, f_1, f_2, \ldots, f_t$ be a sequence of feasible partial embeddings such that for each $i$, $f_{i+1}$ succeeds $f_i$ and $L(f_0) = R(f_t) = \emptyset$. Let $S_i$ be the domain of $f_i$. First we show that for every vertex $v$ there is an index $i$ such that $v \in S_i$. If $v \notin S_0$, then $v \in R(f_0)$. Let $k$ be the largest integer such that $v \in R(f_k)$. Because $R(f_t) = \emptyset$, we have that $k < t$. Thus, $v \in R(f_k) \setminus R(f_{k+1})$. By Lemma 2.7, $R(f_k) \setminus R(f_{k+1}) \subseteq S_{f_{k+1}}^{d+1}$ which implies that $v \in S_{k+1}$.

We claim that for every $v \in S_i \cap S_j$, $f_i(v) + i = f_j(v) + j$. Indeed, let $k$ be the smallest integer such that $v \in S_k$. Let $k' = \min\{t, f_k(v) + k + d + 1\}$. For every $i$ and $j$, such that $k \le i, j \le k'$, we have $f_i(v) + i = f_j(v) + j$. Furthermore, if $k' < t$, then $v \in L(f_{k'+1})$ and thus by Lemma 2.7, $v \in L(f_{k''})$ for every $k' < k'' \le t$. Since $k$ is the smallest integer such that $v \in S_k$, we have that if $v \in S_i \cap S_j$, then $f_i(v) + i = f_j(v) + j$.

From the previous two paragraphs, we conclude that there is a function $f$ such that for every $v \in S_i$, $f(v) = f_i(v) + i$. It remains to prove that $f$ is a distortion $d$ embedding of $G$ into the line. We say that a pair of vertices $u$ and $v$ are *in conflict* if either $|f(u) - f(v)| < D(u,v)$, or if $|f(u) - f(v)| > d \cdot D(u,v)$. Let us note that if no pair of vertices are in conflict, then $f$ is a distortion $d$ embedding of $G$. We prove that no two vertices in $S_i \cup L(f_i)$ are in conflict by induction on $i$. For $i = 0$ this is true as $f_0$ is a feasible partial embedding. Assume now that the statement is true for every $i < k$.

If $S_{f_k}^{d+1}$ is empty, then the statement trivially holds for $k$. Otherwise, for some vertex $v$, $S_{f_k}^{d+1} = \{v\}$. To complete the proof, it is sufficient to show that $v$ is not in conflict with any other vertex $u$ in $S_k \cup L(f_k)$. If $u$ is in $S_k$, $u$ and $v$ are not in conflict because $f_k$ is a feasible partial embedding. If $u$ is not in $S_k$, then $u$ is in $L(f_k)$ and every shortest path from $u$ to $v$ in $G$ must contain a vertex $w \in S_k^L$. Since $f(u) \le f(w) \le f(v)$, we have that $|f(v) - f(u)| = f(v) - f(w) + f(w) - f(u) \ge D(v,w) + D(w,u) = D(v,u)$. Therefore, $|f(v) - f(u)| = f(v) - f(w) + f(w) - f(u) \le d \cdot D(v,w) + d \cdot D(w,u) = d \cdot D(v,u)$. Thus no pair of vertices in $S_i \cup L(f_i)$ are in conflict for every $i \le t$. However, for $i = t$, $S_i \cup L(f_i) = V(G)$ and we conclude that no pair of vertices are in conflict. □

For a vertex $v$ of a graph $G$ and integer $r \ge 0$ we denote the ball of radius $r$ centered in $v$, which is the set of vertices at distance at most $r$ in $G$, by $B(v,r)$. The *local density* of a graph $G$ is $\delta = \max_{v \in V(G), r > 0} \frac{|B(v,r)| - 1}{2r}$. We will apply the following well known lower bound on distortion.

**Lemma 2.9** ([1]). [Local Density] *Let $G$ be a graph that can be embedded into the line with distortion $d$. Then $d$ is at least the local density $\delta$ of $G$.*

Applying Lemma 2.9 we can bound the number of possible feasible partial embeddings. Observe that each feasible partial embedding $f$ can be represented as a number $1 \le t \le d$ and a sequence of vertices $v_0 v_1 \ldots v_q$ such that $t + \sum_{i=1}^{q} D(v_{i-1}, v_i) \le 2d + 1$ and $D(v_{i-1}, v_i) \le d$ for every $i \ge 1$. This is done by simply saying that the domain $S$ of $f$ is the set $\{v_0, v_1, \ldots, v_q\}$ and that $f(v_a) = -(d+1) + t + \sum_{i=1}^{a} D(v_{i-1}, v_i)$. Let $\mathcal{N}(x)$ be the maximum number of sequences $v_0 v_1 \ldots v_q$ such that $\sum_{i=1}^{q} D(v_{i-1}, v_i) = x$, where maximum is taken over all $v_0 \in V(G)$. For any negative number $x$, $\mathcal{N}(x) = 0$.

**Lemma 2.10.** *For $x \in \mathbb{Z}$, $\mathcal{N}(x) \le (2d+1)^x$.*

*Proof.* We prove the Lemma by induction on $x$. For $x \le 0$, the statement is trivially true. Suppose that the inequality holds for every $x' < x$. For a vertex $v_0$, let $\mathcal{S}$ be the set of all vertex sequences $v_0 v_1 \ldots v_q$ starting with $v_0$ with the property that $\sum_{i=1}^{q} D(v_{i-1}, v_i) = x$. For $i \in \{1, \ldots, x\}$, let $\mathcal{S}_i$ be the set of sequences in $\mathcal{S}$ such that $D(v_0, v_1) = i$. Let $C(v_0, i) = |B(v_0, i) \setminus B(v_0, i-1)|$. Then $|\mathcal{S}_i| \le C(v_0, i) \cdot \mathcal{N}(x-i)$ and $|\mathcal{S}| = \sum_{i=1}^{x} |\mathcal{S}_i| \le \sum_{i=1}^{x} C(v_0, i) \cdot \mathcal{N}(x-i)$. By the induction assumption, $\sum_{i=1}^{x} C(v_0, i) \cdot \mathcal{N}(x-i) \le \sum_{i=1}^{x} C(v_0, i) \cdot (2d+1)^{x-i}$. Furthermore, by Lemma 2.9, we have that $\sum_{j=1}^{i} C(v_0, i) \le 2di$ for every $i$. Because $(2d+1)^y$ is a convex function of $y$, it follows that the sum $\sum_{i=1}^{x} C(v_0, i) \cdot (2d+1)^{x-i}$ subject to the constraints $\sum_{j=1}^{i} C(v_0, j) \le 2di$, $1 \le i \le x$, is maximized when each of $C(v_0, i) = 2d$. In this case $\sum_{i=1}^{x} C(v_0, i) \cdot (2d+1)^{x-i} \le 2d \cdot \sum_{i=1}^{x} (2d+1)^{x-i}$ which is a geometric sequence with sum upper bounded by $2d \cdot (2d+1)^x \cdot \sum_{i=1}^{\infty} (2d+1)^{-i} = (2d+1)^x$. Since this holds for each choice of $v_0$, the inequality holds also for $x$. □

**Corollary 2.11.** *For a graph $G$ with local density at most $d$ the number of possible feasible partial embeddings of subsets of $V(G)$ is at most $O(n(2d+1)^{2d})$.*

*Proof.* By discussions preceding Lemma 2.10, for each fixed first vertex $v_0$ and each value of $t$, there are at most $\mathcal{N}(2d+1-t)$ feasible partial embeddings that map $v_0$ to $-(d+1) + t$ Thus the number of feasible partial embeddings is at most $\sum_{t=1}^{d} n \mathcal{N}(2d+1-t)$. By Lemma 2.10, this is at most $n \cdot \sum_{t=1}^{d} (2d+1)^{2d+1-t} \le \frac{3}{2} n (2d+1)^{2d}$. □

Now we are in the position to prove Theorem 1.1.

*Proof.* [**of Theorem 1.1**] The algorithm proceeds as follows. First, check whether $G$ has local density $\delta$ bounded by $d$. Checking the local density of $G$ can be done in time linear in $n$ because if $|E(G)| \ge nd$ we can immediately answer "no". If $\delta > d$, answer "no". Otherwise, we can test whether the conditions of theorem 2.8 apply. That is, we construct a directed graph $\mathcal{D}$ where the vertices are feasible partial embeddings and there is an edge from a partial embedding $f_x$ to a partial embedding $f_y$ if $f_y$ succeeds $f_x$. Checking the conditions of Theorem 2.8, reduces to checking for the existence of a directed path starting in a feasible partial embedding $f_0$ with $L(f_0) = \emptyset$ and ending in a feasible partial embedding $f_t$ with $R(f_t) = \emptyset$. This can be done in linear time in the size of $\mathcal{D}$ by running a depth first search in $\mathcal{D}$. The number of vertices in $\mathcal{D}$ is at most $O(n(2d+1)^{2d})$. Every vertex of $\mathcal{D}$ has at most $O(d^2)$ edges going out of it, as a feasible partial embedding $f_y$ succeeding another feasible partial embedding $f_x$ is completely determined by $f_x$ together with the vertex that $f_y$ maps to $d+1$ (or the fact that $f_y$ does not map anything there). Using prefix-tree-like data structures one can test whether a given partial embedding $f_x$ succeeds another in $O(d^2)$ time. The total running time is then bounded by $O(nd^4(2d+1)^{2d})$. □

## 2.2 Weighted Graph Metrics into the Line parameterized by $d$ and $W$

In the previous section, we gave an FPT algorithm for embedding unweighted graph metrics into the line. Here, we generalize this result to handle metrics generated by weighted graphs. More precisely, let $G = (V, E)$ be a graph with weight function $w : E \to \mathbb{Z}^+ \setminus \{0\}$ and $M = (V(G), D)$ be the weighted shortest path distance metric of $G$. Now we give an outline of an algorithm for embedding $M$ into the line, parameterized by the distortion $d$ and the maximum edge weight $W$, that is, $W = \max_{e \in E}\{w(e)\}$. The definition of a pushing embedding and Observations 2.1 and 2.2 work out even when $G$ is a weighted graph. Once we define the notion of partial embeddings, other notions like feasibility and succession are adapted in an obvious way. Given a graph $G$ and a subset $S \subseteq V(G)$, a *partial embedding* of $S$ is a function $f : S \to \{-(dW+1), \ldots, (dW+1)\}$. We can prove results analogous to Lemma 2.7 and Theorem 2.8 with the new definitions of *partial embeddings, feasibility* and *succession*. Thus, we can give an algorithm for this problem similar to the algorithm presented in Theorem 1.1. The runtime of this algorithm is dominated by the number of different feasible partial embeddings. Let $B_w(v, r)$ denote the set of vertices at *weighted* distance at most $r$ from $v$ and $\delta_w$ be the analogous notion of *weighted local density* of a graph $G$. It is easy to see that if $M$ can be embedded into the line with distortion at most $d$ then $d \geq \delta_w$. This result immediately upper bounds the number of feasible partial embeddings by $n \cdot (dW)^{O(dW)}$. In what follows next we show that the number of feasible partial embeddings actually is bounded by $n \cdot (2d+1)^{2dW}$. Let $\mathcal{N}(x)$ be as in Lemma 2.10. For each fixed first vertex $v_0$ in the partial embedding, and each value of $1 \leq t \leq (2dW+1)$, there are at most $\mathcal{N}(2dW+1-t)$ feasible partial embeddings that map $v_0$ to $-(dW+1)+t$. Thus the number of feasible partial embeddings is at most $\sum_{t=1}^{dW} n \cdot \mathcal{N}(2dW+1-t)$. By Lemma 2.10, this is at most $n \cdot \sum_{t=1}^{dW}(2d+1)^{2dW+1-t} \leq \frac{3}{2}n(2d+1)^{2dW}$.

**Theorem 2.12.** *Given a weighted graph $G$ with maximum edge weight $W$ we can decide whether $M(G)$ can be embedded into the real line with distortion at most $d$ in time $O(n(dW)^4(2d+1)^{2dW})$.*

## 3 Graph Metrics into the Line is Hard for Fixed Rational $d \geq 2$

We complement Theorem 2.12 by proving that deciding whether a given weighted graph metric can be embedded into the line with distortion at most $d$ is NP-complete for every fixed rational $d \geq 2$. Our reduction is from 3-COLORING, one of the classical $NP$-complete problems. On input $G = (V, E)$ to 3-COLORING we show how to construct an edge weighted graph $G' = (V', E')$. For an edge $uv \in E'$, $w(uv)$ will be the weight if the edge $uv$. The weighted shortest path metric $M(G')$ will then be the input to our embedding problem. Let $n = |V|$, $m = |E|$ and $d = \frac{a}{b} \geq 2$ for some integers $a$ and $b$. Let $e_1, e_2, \ldots, e_m$ be an ordering of the edges of $G$, and choose the integers $g = 5a - 1$, $r = 10b$, $q = m(2n+1)$, $L = 10qb$ and $t = abL + 1$. We start constructing $G'$ by making two cliques $C_1$ and $C_2$ of size $t$. Let $C_1 = \{c_1, c_2, \ldots, c_t\}$ and $C_2 = \{c'_1, c'_2, \ldots, c'_t\}$. Let $w(c_ic_j) = w(c'_ic'_j) = \lceil |i - j|/d \rceil$. Now, we make $q - 1$ *separator vertices* and label them $s_1, \ldots, s_{q-1}$. We make $q$ gadgets $T_1, \ldots, T_q$ encoding the edges of $G$. For every edge $e_i = uv$ there are $2n+1$ gadgets, namely $T_{i+mp}$ for every $0 \leq p < 2n+1$. Each such gadget, say $T_{i+mp}$, consists of three vertices, one vertex corresponding to $u$, one vertex corresponding to $v$ and one vertex corresponding to $e_i$. These three vertices form a triangle with edges of weight 1. For every $j$ between 1 and $q$ we connect all vertices of $T_j$ to $s_{j-1}$ and $s_j$ with edges of weight $g$. Whenever this implies that we need to connect something to the non-existing vertices $s_0$ and $s_q$ we connect to $c_t$ and $c'_1$ respectively. Now, for every pair of vertices $x \in T_i$ and $y \in T_j$ that correspond to the same vertex or edge of $G$ we add an edge of weight $r|i-j|$ between $x$ and $y$. Finally, we add an edge with weight $L$ between $c_t$ and $c'_1$. This concludes the construction of $G'$. Figure 1 shows the general structure of the construction. The next lemma essentially shows that if there is an edge $uv \in E'$ then that is the shortest weight path between $u$ and $v$ in $G'$.

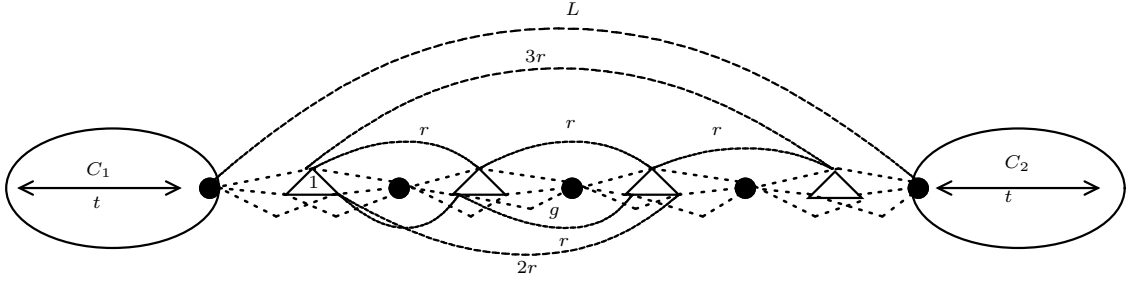

Figure 1: The figure shows the overall structure of the construction. The numbers appearing between $C_1$ and $C_2$ indicate edge weights.

**Lemma 3.1.** [⋆] *For every edge $uv$ in $E'$, $D_{G'}(u,v) = w(uv)$.*

**Lemma 3.2.** [⋆] *If $G$ is 3-colorable then there is an embedding $f$ of $M(G')$ into the line with distortion at most $d$.*

**Lemma 3.3.** *If there is an embedding $f$ of $M(G')$ into the line with distortion at most $d$ then $G$ is 3-colorable.*

*Proof.* Without loss of generality, we assume that $f$ is a pushing embedding (Observation 2.2). Let $\sigma$ be the ordering of the vertices of $G$ imposed by $f$. Now we describe the structure of the ordering $\sigma$. Towards this, we first prove that $\sigma$ orders the vertices of the clique $C_1$ consecutively. That is, if $u$ and $v$ are the leftmost and the rightmost vertex of $C_1$ with respect to the ordering $\sigma$, then there is no vertex $w$ in $V' \setminus C_1$ such that $f(u) < f(w) < f(v)$. We know that $|f(u) - f(v)| \geq |C_1| - 1$. However $c_1 c_t$ is the only edge in $C_1$ satisfying $|f(u) - f(v)| \leq w(uv)d$. Furthermore $w(c_1 c_t)d = |C_1| - 1$ and hence $\sigma$ must order the vertices of $C_1$ consecutively with $c_1$ and $c_t$ as its endpoints. Similarly $\sigma$ must order the vertices of $C_2$ consecutively with $c'_1$ and $c'_t$ as its endpoints. Also, without loss of generality we can assume that $C_1$ appears before $C_2$ in our ordering, because if $C_2$ appears first we can reverse our ordering. Now, if $c_t$ is the leftmost vertex of $C_1$ or $c'_1$ is the rightmost vertex of $C_2$ then the edge $c_t c'_1$ is stretched by a factor more than $d$, as $t > Ld$. Thus, $c_t$ is the rightmost endpoint of $C_1$ and $c'_1$ is the leftmost endpoint of $C_2$. Now, every vertex not in $C_1$ or $C_2$ has to appear in between $C_1$ and $C_2$ because no edge with at least one endpoint outside of $C_1 \cup C_2$ is long enough to stretch over the entire expanse of $C_1$ or $C_2$.

Next, we prove that $\sigma$ orders the vertices as follows $C_1, T_1, s_1, T_2, s_2, T_3, \ldots, T_q, C_2$. To show this, we introduce the notion of gaps. A *gap* between two vertices $u$ and $v$ appearing consecutively in $\sigma$ is simply the interval $[f(u), f(v)]$ on the real line. We say that a gap is *incident* to a vertex $u$ if the vertex $u$ is one of the endpoints of the gap. The size of the gap is $|f(u) - f(v)|$. In the layout, there are $4q - 1$ vertices and $4q$ gaps that appear between $c_t$ and $c'_1$. In the following discussion we will treat $c_t$ and $c'_1$ as separator vertices. Each gap that is incident to two separator vertices, one separator vertex and no separator vertex has size at least $2g$, $g$ and 1 respectively. Let $x_0$, $x_1$ and $x_2$ be the number of gaps incident to 0, 1 and 2 separator vertices respectively.

Then $|f(c_t) - f(c_1')| \geq 2gx_2 + gx_1 + x_0$ and $x_0 = 4q - x_2 - x_1$. Furthermore each separator vertex (except $c_t$ and $c_1'$) is incident to exactly two gaps, while $c_t$ and $c_1'$ are incident to exactly one gap each among the gaps between $c_t$ and $c_1'$. Therefore we have that $x_1 + 2x_2 = 2q$. Substituting $x_1 = 2q - 2x_2$, we get $x_0 = 2q + x_2$ and $|f(c_t) - f(c_1')| \geq 2gq + x_0$. Hence, if $x_2 > 0$ we have $|f(c_t) - f(c_1')| > 2gq + 2q = 2(5a-1)q + 2q = 10aq = 10aqb/b = 10qbd = 10Ld = w(c_tc_1')d$, contradicting that the expansion of $f$ is at most $d$. Thus, $x_2 = 0$, $x_1 = x_0 = 2q$ and hence $|f(c_t) - f(c_1')| \geq 2gq + 2q = w(c_tc_1')d$. Also, if any gap not incident to any separator vertices has size more than 1, or if any of the gaps incident to a separator vertex have size more than $g$ then $|f(c_t) - f(c_1')| > 2gq + 2q = w(c_tc_1')d$, again contradicting that the expansion of $f$ is at most $d$. Finally note that $g > d$ and hence no edge with weight one can ever be stretched over a gap of size $g$. Since the only edges of weight 1 in $G'$ are within a gadget $T_i$ and every edge incident to a separator vertex has weight $g$ this implies that $\sigma$ must order the vertices in the aforementioned order $C_1, T_1, s_1, T_2, s_2, T_3, \ldots, T_q, C_2$. This concludes the description of ordering $\sigma$.

For a vertex $v$ in $V$, if there is a vertex $v'$ in a gadget $T_i$ corresponding to $v$, we look at the position that $v'$ is assigned by $\sigma$ compared to the other vertices of $T_i$. If the relative position of $v'$ given by $\sigma$ with respect to other vertices of $T_i$ is $k \in \{1, 2, 3\}$, then we say that the *color* of $v$ in the gadget $T_i$ is $k$ and denote it by $\chi(v, i)$. In all of these cases we say that $v$ has a color in $T_j$. We prove that for any $i, j$ with $i < j$ and vertex $v \in V$ such that $v$ has a color in both $T_i$ and $T_j$ then $\chi(v, j) \leq \chi(v, i)$. Suppose this is not the case, and let $u'$ and $v'$ be the vertices corresponding to $v$ in gadgets $T_i$ and $T_j$ such that $\chi(v, j) > \chi(v, i)$. Then we know that $|f(v') - f(u')| > (2g+2)|j-i| = (2(5a-1)+2)|j-i| = 10a|j-i|b/b = 10b|j-i|d = r|j-i|d$. However since both $u'$ and $v'$ correspond to $v$ there is an edge of weight $r|j-i|$ between $u$ and $v$ which is stretched more than $d$ by the embedding. Thus we obtain a contradiction which allows us to conclude that $\chi(v, j) \leq \chi(v, i)$. Notice that since a vertex (in a gadget) can have one of three different colors this implies that as we scan the gadgets from $T_1$ to $T_q$ the color of a vertex can change at most twice. Thus, there must be some $0 \leq p < 2n+1$ such that every vertex of $G$ has the same color in all the gadgets it appears in among $T_{1+mp}$ to $T_{m(p+1)}$. Notice that every vertex and every edge of $G$ has a color in at least one of these gadgets. We can now make a coloring $\psi$ of the vertices of $G$. For every vertex $v \in V$, we look at the gadget $T_i$, $1 + mp \leq i \leq m(p+1)$, such that there is a vertex corresponding to $v \in T_i$ and assign $\psi(v) = \chi(v, i)$. All that remains to prove is that $\psi$ is a proper coloring. For every edge $uv \in E$ there is an $i$ between $1+mp$ and $m(p+1)$ such that the edge $uv$ has a color in $T_i$. Then both $u$ and $v$ have colors in $T_i$ and their colors in $T_i$ must be different. Since $\psi(u)$ is equal to $u$'s color in $T_i$ and $\psi(v)$ is equal to $v$'s color in $T_i$ this implies $\psi(u) \neq \psi(v)$ concluding the proof. □

Together with the construction of $G'$ from $G$, Lemmas 3.2 and 3.3 imply Theorem 1.2.

## 4 Embedding Graphs into Trees of Bounded Degree

Given a graph $G$ with shortest path metric $D_G$ and a tree $T$ with maximum degree $\Delta$, having shortest path metric $D_T$, we give an algorithm that decides whether $G$ can be embedded into $T$ with distortion at most $d$ in time $n^2 \cdot |V(T)| \cdot 2^{O((5d)^{\Delta+1} \cdot d)}$. We assume that the tree $T$ is rooted, and we will refer to the root of $T$ as $r(T)$. For a vertex $v$ in the tree, $T_v$ is the subtree of $T$ rooted at $v$, and $C(v)$ is the set of $v$'s children. Finally, for an edge $uv$ of $T$, let $T_u(uv)$ and $T_v(uv)$ be the tree of $T \setminus uv$ that contains $u$ and $v$ respectively. Notice that if $u$ is the parent of $v$ in the tree, then $T_v(uv) = T_v$ and $T_u(uv) = T \setminus V(T_v)$. As in the previous section, we need to define feasible partial embeddings together with the notion of succession. For a vertex $u \in V(T)$ and a subset $S$ of $V(G)$, a *u-partial embedding* is a function $f_u : S \to B(u, d+1)$.

**Definition 4.1.** *For a u-partial embedding $f_u$ of a subset $S \subseteq V(G)$ and a vertex $v \in N(u)$ we define $S[v, f_u] = \{x \in S : f_u(x) \in V(T_v(uv))\}$. Given two integers $i$ and $j$, $0 \leq i \leq j \leq k$,* let $S^{[i,j]}[f_u] = \{x \in S : i \leq D(f_u(x), u) \leq j\}$. Finally, let $S^{[i,j]}[v, f_u] = S^{[i,j]}[f_u] \cap S[v, f_u]$, $S^k[v, f_u] = S^{[k,k]}[v, f_u]$ for $k \geq 1$ and $S^0[f_u] = S^{[0,0]}[f_u]$.

**Definition 4.2.** *For a $u$-partial embedding $f_u$ of a subset $S \subseteq V(G)$ and a vertex $v \in N(u)$ we define $M[v, f_u]$ to be the union of the vertex sets of all connected components of $G \setminus S$ that have neighbors in $S[v, f_u]$.*

**Definition 4.3.** *A $u$-partial embedding $f_u$ of a subset $S$ of $V(G)$ is called feasible if (1) $f_u$ is a non-contracting distortion $d$ embedding of $S$ into $B(u, d+1)$; (2) for any distinct pair $v, w \in N(u)$, $M[v, f_u] \cap M[w, f_u] = \emptyset$; (3) $N(S^0[f_u]) \subseteq S$.*

**Definition 4.4.** *For a feasible $u$-partial embedding $f_u$ of a subset $S_u$ of $V(G)$ and a feasible $v$-partial embedding $f_v$ of a subset $S_v$ of $V(G)$ with $v \in C(u)$ we say that $f_v$ succeeds $f_u$ if (1) $S_u \cap S_v = (S_u^{[0,d]}[f_u] \cup S_u^{d+1}[v, f_u]) = (S_v^{[0,d]}[f_v] \cup S_v^{d+1}[u, f_v])$; (2) for every $x \in S_u \cap S_v$, $f_u(x) = f_v(x)$; (3) $M[v, f_u] = \bigcup_{x \in N(v) \setminus u}(M[x, f_v] \uplus S_u^{d+1}[x, f_v])$; and (4) $M[u, f_v] = \bigcup_{x \in N(u) \setminus v}(M[x, f_u] \uplus S_v^{d+1}[x, f_u])$.*

Suppose we have picked out a subtree $T_v$ for a vertex $v \in V(T)$ and found a non-contracting embedding $f'$ with distortion at most $d$ of a subset $Z$ of $V(G)$ into $T' = T[\bigcup_{u \in V(T_v)} B(u, d+1)]$. We wish to find a non-contracting distortion $d$ embedding of $G$ into $T$ such that for every vertex $u$ with $f(u) \in V(T')$, we have that $u \in Z$ and such that if $u \in Z$ then $f(u) = f'(u)$. At this point, a natural question arises. Can we impose constraints on the restriction of $f$ to $V(T) \setminus V(T_v)$ such that $f$ restricted to $V(T) \setminus V(T_v)$ satisfies these conditions if and only if $f$ is a non-contracting distortion $d$ embedding of $G$ into $T$? One necessary condition is that $f$ restricted to $V(T) \setminus V(T_v)$ must be a non-contracting distortion $d$ embedding of $\{u \in V(G) : f(u) \in V(T) \setminus V(T_v)\}$. We can obtain another condition by applying the definition of feasible $u$-partial embeddings. For each vertex $u$, we can use arguments similar to the ones in Section 2 in order to determine which connected components of $T \setminus V(T_v)$ $f$ must map $u$ to in order to be a non-contracting distortion $d$ embedding of $G$ into $T$.

For the line, these two conditions are both necessary and sufficient. Unfortunately, for the case of bounded degree trees, this is not the case. The reason the conditions are sufficient when we restrict ourselves to the line is that every embedding of a graph metric into the line that is *locally* non-contracting and *locally* expanding by a factor at most $d$, also is *globally* non-contracting and expanding by a factor at most $d$. When we embed into trees of bounded degree, every embedding that is locally expanding by a factor at most $d$, also has this property globally. However, every locally non-contracting embedding need not be *globally* non-contracting. To cope with this issue, we introduce the concept of vertex types. Intuitively, vertices of the same type in $T_v$ are indistinguishable when viewed from $T \setminus V(T_v)$. We show that the set of possible vertex types can be bounded by a function of $d$ and $\Delta$. Then, to complete $f$ from $f'$ we only need to know the restriction of $f'$ to $B(v, d+1)$ and which vertex types appear in $T_v$. Then the amount of information we need to pass on from $f'$ to $f$ is bounded by $n \cdot h(d, \Delta)$. We exploit this fact to give an algorithm for the problem. In the rest of this section, we formalize this intuition.

For a vertex $u \in T$, a neighbor $v$ of $u$ and a feasible $u$-partial embedding $f_u$ of a subset $S$ of $V(G)$ we define a $[v, f_u]$-*type* to be a function $t : S[v, f_u] \to \{\infty, 3d + 2, d, \ldots, -d, -(d+1)\}$ and a $[v, f_u]$-*typelist* to be a set of $[v, f_u]$-types. For an integer $k$ let $\beta(k) = k$ if $k \leq 3d + 2$ and $\beta(k) = \infty$ otherwise.

**Definition 4.5.** *For a vertex $u \in T$ with two neighbors $v$ and $w$, and a feasible $u$-partial embedding $f_u$ of a subset $S$ of $V(G)$ together with a $[v, f_u]$-typelist $\mathcal{L}_1$ and a $[w, f_u]$-typelist $\mathcal{L}_2$ we say that $\mathcal{L}_1$ and $\mathcal{L}_2$ agree if for every type $t_1 \in \mathcal{L}_1$ and $t_2 \in \mathcal{L}_2$ there is a vertex $x \in S[v, f_u]$ and a vertex $y \in S[w, f_u]$ such that $t_1(x) + t_2(y) \geq D_G(x, y)$.*

**Definition 4.6.** *For a vertex $u \in T$, a neighbor $v$ of $u$, a feasible $u$-partial embedding $f_u$ of a subset $S$ of $V(G)$ and a $[v, f_u]$-typelist $\mathcal{L}$ we say that $\mathcal{L}$ is* compatible *with $S[v, f_u]$ if for every vertex $x$ in $S[v, f_u]$ there is a type $t \in \mathcal{L}$ such that for every $y \in S[v, f_u]$, $D_T(f_u(x), u) - D_G(x, y) = t(y)$.*

**Definition 4.7.** *A* feasible $u$-state *is a feasible partial embedding $f_u$ of a subset $S$ of $V(G)$ together with a $[v, f_u]$-typelist $\mathcal{L}[v, f_u]$ for every $v \in N(u)$ such that the following conditions are satisfied: (1) $\mathcal{L}[v, f_u]$ is compatible with $S[v, f_u]$ for every $v \in N(u)$; and (2) For every pair of distinct vertices $x$ and $y$ in $N(u)$, $\mathcal{L}[x, f_u]$ agrees with $\mathcal{L}[y, f_u]$.*

**Definition 4.8.** *Let $u \in V(T)$, $v \in C(u)$. Let $\mathcal{X}_u$ be a feasible $u$-state and $\mathcal{X}_v$ be a feasible $v$-state. We say that $\mathcal{X}_v$* succeeds *$\mathcal{X}_u$ if*

1. *$f_v$ succeeds $f_u$;*
2. *For every $w \in (N(v) \setminus u)$ and a type $t_1 \in \mathcal{L}[w, f_v]$ there is a type $t_2 \in \mathcal{L}[v, f_u]$ such that*
   
   (a) *For every node $x \in S[v, f_u] \cap S[w, f_v]$, $t_2(x) = \beta(t_1(x) + 1)$;*
   (b) *For every node $x \in (S[v, f_u] \setminus S[w, f_v])$, $t_2(x) = \beta(\max_{y \in S[w, f_v]}(t_1(y) + 1 - D_G(x, y)))$.*

3. *For every $w \in (N(u) \setminus v)$ and a type $t_1 \in \mathcal{L}[w, f_u]$ there is a type $t_2 \in \mathcal{L}[u, f_v]$ such that*
   
   (a) *For every node $x \in S[u, f_v] \cap S[w, f_u]$, $t_2(x) = \beta(t_1(x) + 1)$;*
   (b) *For every node $x \in (S[u, f_v] \setminus S[w, f_u])$, $t_2(x) = \beta(\max_{y \in S[w, f_u]}(t_1(y) + 1 - D_G(x, y)))$.*

The main result of this section relies on the next two lemmas.

**Lemma 4.9.** [⋆] *If there is a distortion $d$ embedding $F$ of $G$ into $T$ then, for every vertex $u$ of $V(T)$ there is a feasible $u$-state $\mathcal{X}_u$ such that for every vertex $v \in V(T), w \in C(v)$, $\mathcal{X}_w$ succeeds $\mathcal{X}_v$.*

**Lemma 4.10.** [⋆] *If there is a feasible $u$-state $\mathcal{X}_u$ for every vertex $u$ of $V(T)$ such that for every vertex $v \in V(T), w \in C(v)$, $\mathcal{X}_w$ succeeds $\mathcal{X}_v$ then there is a distortion $d$ embedding $F$ of $G$ into $T$.*

**Theorem 1.3** [⋆] *Given a graph $G$, tree $T$ with maximum degree $\Delta$ and an integer $d$ we can decide whether $G$ can be embedded into $T$ with distortion at most $d$ in time $n^2 \cdot |V(T)| \cdot 2^{O((5d)^{\Delta+1} \cdot d)}$.*

## 5 Concluding Remarks and Open Problems

In this paper we described FPT algorithms for embedding unweighted graph metrics into a tree metric for a tree of maximum degree $\Delta$, parameterized by $(\Delta, d)$ where $d$ is the distortion. For the case when the host metric is the line, we generalized our result and showed that embedding weighted graph metrics into the line is FPT parameterized by distortion $d$ and maximum edge weight $W$. A similar generalization can also be obtained for embedding weighted graph metrics into weighted bounded degree tree metrics, parameterized by $d$, $\Delta$ and $W$ where $W$ is the maximum edge weight in the input graph. We postpone the details for the full version of the paper. Our hardness result that embedding a weighted metric into the line is NP-hard for every fixed distortion $d \geq 2$ showed that our algorithms qualitatively are the best possible.

We believe that our results will lead to further investigation of the combinatorially challenging field of low distortion embeddings within the framework of multivariate algorithmics. We conclude with two concrete interesting open problems:

- What is the parameterized complexity of embedding unweighted graph metrics into unbounded degree trees, parameterized by distortion $d$?
- What is the parameterized complexity of embedding unweighted graph metrics into target metrics that are minimum distance metrics of cycles, parameterized by $d$?

# 6 Appendix

**Proof of Observation 2.1:** *If $f(u) < f(v) < f(w)$ and $u$ pushes $w$, then $u$ pushes $v$ and $v$ pushes $w$.*

*Proof.* By the triangle inequality, $D(u,w) \leq D(u,v) + D(v,w)$. Since $u$ pushes $w$ and because $f$ is non-contracting, we have that $D(u,w) = f(w) - f(u) = (f(w) - f(v)) + (f(v) - f(u)) \geq D(u,v) + D(v,w) = D(u,w)$. Since $f(w) - f(v) \geq D(v,w)$ and $f(v) - f(u) \geq D(u,v)$ we have $f(w) - f(v) = D(v,w)$ and $f(v) - f(u) = D(u,v)$. □

**Proof of Observation 2.2:** *If $G$ can be embedded into the line with distortion $d$, then there is a pushing embedding of $G$ into the line with distortion $d$. Furthermore, every pushing embedding of $G$ into the line is non-contracting.*

*Proof.* Among all embeddings of $G$ into the line with distortion $d$, let us choose $f$ such that

$$\sum_{2 \leq i \leq n} |f(v_i) - f(v_{i-1})| \text{ is minimized.}$$

We claim that $f$ is pushing. Indeed, if $f$ is not pushing, then there is a minimum integer $q \geq 1$ such that $v_q$ does not push $v_{q+1}$. By Observation 2.1, for every $p \leq q$ and $r \geq q+1$, $D(v_p, v_r) > f(v_r) - f(v_p)$. But then embedding $f'$, $f'(v_i) = f(v_i)$ for $i \leq q$ and $f'(v_i) = f(v_i) - 1$ for $i > q$, is non-contracting embedding of distortion $d$, which is a contradiction to the choice of $f$.

To prove that every pushing embedding of $G$ into the line is non-contracting, we observe that for each $b > a \geq 1$, $f(v_b) - f(v_a) = \sum_{i=a+1}^{b} f(v_i) - f(v_{i-1}) = \sum_{i=a+1}^{b} D(v_i, v_{i-1}) \geq D(v_a, v_b)$. □

**Proof of Lemma 2.7:** *For every pair of feasible partial embeddings $f$ and $g$ of subsets $S_f$ and $S_g$ of $V(G)$ such that $g$ succeeds $f$, we have $R(f) = R(g) \cup S_g^{d+1}$ and $L(g) = L(f) \cup S_f^{-(d+1)}$.*

*Proof.* Let us prove that $R(f) = R(g) \cup S_g^{d+1}$. (The proof of $L(g) = L(f) \cup S_f^{-(d+1)}$ is similar.) Because $g$ succeeds $f$, we have that $S_g^{d+1} \subseteq R(f)$. Let $C$ be the vertex set of a connected component of $G \setminus S_g$ such that $C \subseteq R(g)$. As $S_f^{-(d+1)} \subseteq L(g)$, the subgraph $G[C]$ induced by $C$, is a connected component of $G \setminus (S_f \cup S_g)$. If $C$ contains a neighbor of $S_g^{d+1}$, then $C \subseteq R(f)$ (this is because $S_g^{d+1} \subseteq R(f)$ and $C$ and $S_g^{d+1}$ are in the same connected component of $G \setminus S_f$). On the other hand, if $C$ contains no neighbor of $S_g^{(d+1)}$, then, as $C \subseteq R(g)$, $C$ has a neighbor in $S_g^{(1,d)} \subseteq S_f^R$. Therefore $C \subseteq R(f)$, which in turn implies that $R(g) \subseteq R(f)$. Thus, we have proved that $R(g) \cup S_g^{d+1} \subseteq R(f)$.

Let us now show that $R(f) \subseteq R(g) \cup S_g^{d+1}$. Let $C$ be the vertex set of a connected component of $G \setminus S_f$ such that $C \subseteq R(f)$. $C$ contains no neighbors of $S_f^{-(d+1)}$ thus $C$ is a connected component of $G \setminus S_g^{(-(d+1),d)}$. If $C$ does not contain $S_g^{d+1}$, then $C$ is a connected component of $G \setminus S_g$. Furthermore, as $C \subseteq R(f)$, $C$ has a neighbor in $S_f^{(1,d+1)} \subseteq S_g^{(0,d)}$. As $S_g^0$ has no neighbors outside of $S_g$, $C$ has a neighbor in $S_g^R$ implying $C \subseteq R(g)$. On the other hand, if $C$ contains $S_g^{d+1}$, then every connected component $C'$ of $G[C] \setminus S_g^{d+1}$ is a connected component of $G \setminus S_g$ that has a neighbor in $S_g^{d+1} \subseteq S_g^R$. This concludes the proof that $R(f) \subseteq R(g) \cup S_g^d$, implying $R(f) = R(g) \cup S_g^{d+1}$. □

**Proof of Lemma 3.1:** *For every edge $uv$ in $E'$, $D_{G'}(u,v) = w(uv)$.*

*Proof.* Clearly, $D_{G'}(u,v) \leq w(uv)$ for every edge $uv$, so it is sufficient to prove $D_{G'}(u,v) \geq w(uv)$. If $w(uv) = 1$ then $D_{G'}(u,v) \geq w(uv)$, so suppose $w(uv) > 1$. In this case $uv$ either has

- both endpoints in $C_1$ or $C_2$, or
- is the edge $c_t c'_1$, or
- is an edge from $c_t$ to a vertex in $T_1$, or
- an edge from $c'_1$ to a vertex in $T_q$, or
- an edge incident to a separator vertex or
- an edge between a vertex in a gadget $T_i$ and a vertex in a gadget $T_j$.

If both $u$ and $v$ lie inside $C_1$ every shortest $u - v$ path lies entirely within $C_1$. For every $w$ in $C_1$ we have that $w(u,v) \leq w(u,w) + w(w,v)$ so $w(uv) \leq D_{G'}(u,v)$. Similarly, if both $u$ and $v$ lie inside $C_2$ then $w(uv) \leq D_{G'}(u,v)$. If $uv$ is incident to a separator vertex then $w(uv) = g$ and $D_{G'}(uv) \geq g$ because every separator vertex is only incident to edges with weight $g$. If $uv$ is an edge from $c_t$ to a vertex in $T_1$ or an edge from $c'_1$ to a vertex in $T_q$ then $w(uv) = g$ and $D_{G'}(uv) \geq g$ because every edge with one endpoint inside $C_1 \cup C_2$ and one endpoint outside of $C_1 \cup C_2$ has weight exactly $g$.

Now, if $uv$ is an edge between a vertex in a gadget $T_i$ and a vertex in a gadget $T_j$, $w(uv) = r|i - j|$. Observe that a path from $u$ to $v$ with length smaller than $r|i - j|$ can never use the edge $c_t c'_1$ and thus will never visit the set $C_1 \cup C_2$. We prove that the distance between $u$ and $v$ is at least $r|i - j|$ by induction on $|i - j|$. If $|i - j| = 1$ then any path containing an edge with one endpoint in $T_{i'}$ and another in $T_{j'}$ with $i' \neq j'$ will have length at least $r$. Any path from $u$ to $v$ that does not contain any such edges must contain at least one separator vertex as an in intermediate vertex and thus have length at least $2g = 20a - 2 > 10b = r$. We now suppose that the induction hypothesis is true whenever $|i - j| < z$ and show that it also must hold when $|i - j| = z$. If a path $P$ from $u$ to $v$ contains a vertex $u'$ from a gadget $T_{i'}$ with $i' \neq i$, $i' \neq j$ and $|i' - i| + |j - i'| = |i - j|$ then the induction hypothesis implies that the length of $P$ is at least $|i' - i|r + |j - i'|r = |i - j|r$. If $P$ contains no such vertices as intermediate vertices then $P$ must contain at least one edge with one endpoint in $T_{i'}$ and another in $T_{j'}$ such that $|i' - j'| \geq |i - j|$. In this case the length of $P$ is at least $|i' - j'|r \geq |i - j|r$, concluding the proof that the distance between a vertex $u$ in a gadget $T_i$ and a vertex $v$ in a gadget $T_j$ is at least $r|i - j|$.

It remains to show that $D_{G'}(c_t c'_1) > L$. If a shortest path $P$ from $c_t$ to $c'_1$ avoids the edge $c_t c'_1$, the first vertex in $P$ after $c_t$ must be a vertex $u$ in $T_1$, and the last vertex in $P$ before $c'_1$ must be a vertex $v$ in $T_q$. Thus, by the discussion in the previous paragraph, the length of $P$ is at least $2g + (q - 1)r \geq qr = 10qb = L$, concluding the proof. □

**Proof of Lemma 3.2:** If $G$ is 3-colorable then there is an embedding $f$ of $M(G')$ into the line with distortion at most $d$.

*Proof.* Let $\psi : V(G) \to \{1, 2, 3\}$ be a proper 3-coloring of the vertices of $G$. We extend $\psi$ to also color the edges, by defining $\psi(uv) = \{1, 2, 3\} \setminus \{\psi(u), \psi(v)\}$ for every edge $uv \in E(G)$, that is every edge gets a color different from its two endpoints. We give an ordering of the vertices of $G'$, and the embedding $f$ of $G'$ into the line is the pushing embedding imposed by this ordering. We order the vertices of $G'$ as follows: $C_1, T_1, s_1, T_2, s_2, \ldots, T_q, C_2$. Here, the vertices inside $C_1$ and $C_2$ are ordered like $\{c_1, \ldots, c_t\}$ and $\{c'_1, \ldots, c'_t\}$ respectively and the vertices inside each gadget $T_i$ are ordered by color. That is, if $T_i$ corresponds to an edge $e = uv$ and contains the vertices $u'$, $v'$ and $e'$ corresponding to $u$, $v$ and $e$ respectively, we sort $u$, $v$ and $e$ in increasing order by $\psi$ and use the corresponding order imposed by this for the vertices in $T_i$.

Observation 2.2 implies that the pushing embedding $f$ is non-contracting. Thus, it suffices to show that the expansion of $f$ is at most $d$. Because of Lemma 3.1 it suffices to show that $|f(u) - f(v)| \leq w(uv)d$ for every edge $uv \in E'$. For edges with both endpoints in $C_1$ or both endpoints in $C_2$ this inequality holds. For an edge $uv$ between a separator vertex and a vertex in a gadget $T_i$ we have $|f(u) - f(v)| \leq g + 2 \leq dg = dw(uv)$. For an edge $uv$ between two vertices in the same gadget $T_i$ we have $|f(u) - f(v)| \leq 2 \leq dw(uv)$. Now, for the edge $c_t c'_1$,

$|f(c'_1) - f(c_t)| = (2g + 2)q = 10aq = 10qba/b = Ld = w(c_t c'_1)d$. Similarly, any edge $uv$ with one endpoint in $T_i$ and the other in $T_j$ for $i \neq j$ has the property that $u$ and $v$ correspond to the same vertex (or edge) of $G$ and thus are given the same color by $\psi$. Hence $|f(c'_1) - f(c_t)| = (2g + 2)|i - j| = 10a|i - j| = 10b|i - j|a/b = r|i - j|d = w(uv)d$. As all edges of $G'$ now are accounted for, this means that the expansion of $f$ is at most $d$. □

**Proof of Lemma 4.9** *If there is a distortion $d$ embedding $F$ of $G$ into $T$ then, for every vertex $u$ of $V(T)$ there is a feasible $u$-state $\mathcal{X}_u$ such that for every vertex $v \in V(T), w \in C(v), \mathcal{X}_w$ succeeds $\mathcal{X}_v$.*

*Proof.* We start by giving a feasible $u$-partial embedding $f_u$ for each vertex of the tree. Recall that a feasible $u$-state contains a feasible $u$-partial embedding $f_u$ of a subset $S_u$ of $V(G)$. For a vertex $u \in V(T)$ we define $f_u$ to be the restriction of $F$ to $B(u, d + 1)$. It is easy to see that $f_u$ indeed is a feasible partial embedding for every $u$ and that for every vertex $v \in V(T), w \in C(v)$, $f_w$ succeeds $f_v$.

Now, for every vertex $u \in V(T)$ and $v \in N(v)$ we give a typelist $\mathcal{L}[v, f_u]$. For every vertex $x \in (S[v, f_u] \cup M[v, f_u])$ we add a $[v, f_u]$-type $t_x[v, f_u]$ to $\mathcal{L}[v, f_u]$. For a vertex $y \in S[v, f_u]$, $t_x[v, f_u](y) = \beta(D_T(F(x), u) - D_G(x, y))$. Notice that since $y \in B(u, d + 1)$ and $F$ is non-contracting, by the triangle inequality it follows that $t_x(y) \geq -(d+1)$ and thus $t_x$ is a $[v, f_u]$-type. Furthermore, for every $u \in V(T)$, $\mathcal{L}[v, f_u]$ is compatible with $S[v, f_u]$ because for every $x$ and $y$ in $S[v, f_u]$ we have that $t_x[v, f_u](y) = \beta(D_T(F(x), u) - D_G(x, y))$. In order to show that each state $\mathcal{X}_u$ is a feasible $u$-state it remains to show that for every vertex $u \in V(T)$ and every pair of distinct vertices $v$ and $w$ in $N(u)$, $\mathcal{L}[v, f_u]$ agrees with $\mathcal{L}[w, f_u]$. Assume for contradiction that there is a type $t_a[v, f_u] \in \mathcal{L}[v, f_u]$ and a type $t_b[w, f_u] \in \mathcal{L}[w, f_u]$ such that $t_a[v, f_u](x) + t_b[w, f_u](y) < D_G(x, y)$ for every $x \in S[v, f_u]$ and $y \in S[w, f_u]$. Let $x' \in S[v, f_u]$ and $y' \in S[w, f_u]$ be the pair of vertices that maximizes $t_a[v, f_u](x') + t_b[w, f_u](y') - D_G(x', y')$. There is a vertex $a \in (S[v, f_u] \cup M[v, f_u])$ and a vertex $b \in (S[w, f_u] \cup M[w, f_u])$ such that $\beta(D_T(f(a), u) - D_G(a, x)) = t_a[v, f_u](x)$ for every $x \in S[v, f_u]$ and $\beta(D_T(F(b), u) - D_G(b, y)) = t_b[w, f_u](y)$ for every $y \in S[w, f_u]$. This yields $D_T(F(a), u) - D_G(a, x') + D_T(F(b), u) - D_G(b, y') = (D_T(F(a), u) + D_T(F(b), u)) - (D_G(a, x') + D_G(b, y')) < D_G(x', y')$. Now, $(D_T(F(a), u) + D_T(F(b), u)) = D_T(F(a), F(b))$ since $u$ lies on the unique $f(a)$-$f(b)$ path in $T$. Also, Since $x'$ and $y'$ are the pair that maximize $t_a[v, f_u](x') + t_b[w, f_u](y') - D_G(x', y')$ and every shortest $x'$-$y'$ path in $G$ must pass both through $S[v, f_u]$ and $S[w, f_u]$, we conclude that $(D_G(a, x') + D_G(b, y') + D_G(x', y')) = D_G(a, b)$. However this implies $D_T(F(a), F(b)) < D_G(a, b)$ contradicting that $F$ is non-contracting.

It remains to prove that for every vertex $u \in T, v \in N(u), w \in (N(v) \setminus u)$ and type $t_1 \in \mathcal{L}[w, f_v]$ there is a type $t_2 \in \mathcal{L}[v, f_u]$ such that

1. for every node $x$ in $S[v, f_u] \cap S[w, f_v]$, $t_2(x) = \beta(t_1(x) + 1)$;

2. for every node $x$ in $S[v, f_u] \setminus S[w, f_v]$, $t_2(x) = \beta(\max_{y \in S[w, f_v]}(t_1(y) + 1 - D_G(x, y)))$.

Let $t_a[w, f_v] \in \mathcal{L}[w, f_v]$, and let $a$ be the vertex of $S[w, f_v] \cup M[w, f_v]$ such that for every $x$ in $S[w, f_v]$, $t_a[w, f_v](x) = \beta(D_T(F(a), v) - D_G(a, x))$. Now, $S[w, f_v] \cup M[w, f_v] \subseteq S[v, f_u] \cup M[v, f_u]$ so $a \in (S[v, f_u] \cup M[v, f_u])$. Let $t'_a[v, f_u]$ be the type in $\mathcal{L}[v, f_u]$ so that for every $x'$ in $S[v, f_u]$, $t'_a[v, f_u](x') = \beta(D_T(F(a), u) - D_G(a, x'))$. As $\beta(D_T(F(a), u)) = \beta(D_T(F(a), v) + 1)$ it is easy to see that for every node $x \in (S[v, f_u] \cap S[w, f_v])$, $t'_a[v, f_u](x) = \beta(t_a[w, f_v](x) + 1)$. Finally, observe that for a vertex $x \in (S[v, f_u] \setminus S[w, f_v])$ every $a$-$x$ path in $G$ must pass through $S[w, f_v]$. Thus $D_G(a, x) = \min_{y \in S[w, f_v]} D_G(a, y) + D_G(x, y)$ and so $t'_a[v, f_u](x) = \beta(\max_{y \in S[w, f_v]}(t_a[w, f_v](y) + 1 - D_G(x, y)))$. This concludes the proof. □

**Proof of Lemma 4.10:** *If there is a feasible $u$-state $\mathcal{X}_u$ for every vertex $u$ of $V(T)$ such that for every vertex $v \in V(T), w \in C(v), \mathcal{X}_w$ succeeds $\mathcal{X}_v$ then there is a distortion $d$ embedding $F$ of $G$ into $T$.*

*Proof.* For every vertex $u$, let $f_u$ be the feasible $u$-partial embedding of the subset $S_u \subseteq V(G)$ in $\mathcal{X}_u$. We prove the lemma by proving a series of claims

**Claim 6.1.** *For every vertex $x \in V(G)$ there is a $u \in V(T)$ such that $x \in S_u$.*

*Proof.* If $x \in S_{r(T)}$ we are done, so assume that $x \notin S_{r(T)}$. This means that $x \in \bigcup_{v \in C(r(T))} M[v, f_{r(T)}]$. Let $v_1 \in C(r(T))$ be a vertex such that $x \in M[v_1, f_{r(T)}]$. Observe that the choice of $v_1$ implies that $x \notin M[r(T), f_{v_1}]$. Now, if $x \in S_{v_1}$ we are done, otherwise $x \in \bigcup_{v \in C(v_1)} M[v, f_{v_1}]$. Let $v_2$ be the vertex in $C(v_1)$ so that $x \in M[v_2, f_{v_1}]$. Again, the choice of $v_1$ implies that $x \notin M[v_1, f_{v_2}]$. If $x \in S_{v_2}$ we are done, otherwise we can select $v_3$, $v_4$ and so on until we select a leaf $v_q$. The choice of $v_q$ implies that $x \in S_{v_q} \cup \bigcup_{v \in C(v_q)} M[v, f_{v_q}] = S_{v_q}$. □

**Claim 6.2.** *For every vertex $x \in V(G)$, the set $\{u \in V(T) : x \in S_u\}$ induces a connected subtree of $T$.*

*Proof.* Suppose for contradiction that this is not the case. Then there is a pair of vertices $u, v \in V(T)$ such that $x \in S_u$, $x \in S_v$, $uv \notin E(T)$ and for every $w$ on the $u$-$v$-path in $T$, $x \notin S_w$. Let $w'$ and $w''$ be the predecessor and successor of $w$ on the $u$-$v$ path respectively. By the properties of succession of feasible $u$-partial embeddings both $M[w', f_w]$ and $M[w'', f_w]$ must contain $x$. This contradicts that $f_w$ is a feasible partial embedding. □

From Claim 6.2 together with property 2 of succession feasible $u$-partial embeddings it is clear that for every pair of vertices $u$ and $v$ in $V(T)$ such that $x \in S_u$ and $x \in S_v$, $f_u(x) = f_v(x)$. We can therefore define a function $F : V(G) \to V(T)$ such that for every $x \in V(G)$ and $u \in V(T)$ it holds that if $x \in S_u$ then $F(x) = f_u(x)$. This property also guarantees $F$ maps distinct vertices of $G$ onto distinct vertices of $T$. In the rest of the proof of Lemma 4.10 we will prove that the expansion of $F$ is at most $d$ and that $F$ is non-contracting.

**Claim 6.3.** *The expansion of $F$ is at most $d$.*

*Proof.* It suffices to prove that $F$ expands every edge of $G$ by at most a factor of $d$. Let $xy \in E(G)$. Let $u = F(x)$. By the property 3 of feasible $u$-partial embeddings $y \in S_u$. Furthermore, since $f_u$ is a feasible $u$-partial embedding, $D_T(F(x), F(y)) = D_T(f_u(x), f_u(y)) \leq d$ which completes the proof □

We now proceed to prove that $F$ is non-contracting.

**Claim 6.4.** *For every path $P = v_1 v_2 \ldots v_k$ in $T$, with $v_1 = u$ and $v_k = w$ the following must apply.*

1. *$F$ restricted to $\bigcup_{v_i \in P} S_{v_i}$ is non-contracting.*

2. *For every vertex $x \in S_w$ one of the following two conditions must hold.*

   (a) *either there is a $v_j \in P$, $y \in S_{v_j}$ such that $D_T(F(x), v_j) - D_G(x, y) > 3d + 2$*

   (b) *or there is a type $t_x[v_2, f_u] \in \mathcal{L}[v_2, f_u]$ such that for every $y \in S[v, f_u]$, $t_x[v_2, f_u](y) = D_T(F(x), u) - D_G(x, y)$.*

*Proof.* We prove the claim by induction on $k$. If $k = 1$ then (1) is true because $f_u$ is a feasible partial embedding and (2b) holds because of the compatibility constraints of feasible $u$-states.

For $k = 2$, we first prove that (2b) holds for every $x \in S_w$. If $x \in S_u$ then (2b) holds because of the compatibility constraints of feasible $u$-states. Therefore, consider a vertex $x \in S_w \setminus S_u$. Then $x \in S_w^{d+1}[w', f_w]$ for a $w' \in (N(w) \setminus u)$. By compatibility, there is a type $t_1[w', f_w] \in \mathcal{L}[w', f_w]$ such that for every $y$ in $S_w$, $t_1[w', f_w](y) = D_T(F(x), w) - D_G(x, y)$. By the properties of succession of feasible $u$-states, there is a type $t_2[w, f_u] \in \mathcal{L}[w, f_u]$ such that

1. For every node $y$ in $S[w, f_u] \cap S[w', f_w]$,

$$\begin{aligned}
t_2[w, f_u](y) &= \beta(t_1[w', f_w](y) + 1) \\
&= t_1[w', f_w](y) + 1 \\
&= D_T(F(x), w) - D_G(x, y) + 1 \\
&= D_T(F(x), u) - D_G(x, y).
\end{aligned}$$

2. For every node $y$ in $S[w, f_u] \setminus S[w', f_w]$,

$$\begin{aligned}
t_2[w, f_u](y) &= \beta \left( \max_{z \in S[w', f_w]} (t_1[w', f_w](z) + 1 - D_G(z, y)) \right) \\
&= \max_{z \in S[w', f_w]} (t_1[w', f_w](z) + 1 - D_G(z, y)) \\
&= \max_{z \in S[w', f_w]} (D_T(F(x), w) - D_G(x, z) + 1 - D_G(z, y)) \\
&= D_T(F(x), u) - D_G(x, y).
\end{aligned}$$

Thus (2b) holds for every $x$ in $S_w$. Using this fact we can now prove (1). Observe that it is sufficient to prove that $F$ does not contract any vertex $y \in (S_u \setminus S_w)$ and $x \in (S_w \setminus S_u)$. Let $u'$ be the neighbor of $u$ such that $y \in S[u', f_u]$. By (2b) there is a type $t_x[w, f_u]$ such that for every $z \in S[w, f_u]$, $t_x[w, f_u](z) = D_T(F(x), u) - D_G(x, z)$. By the properties of feasible $u$-states there is a type $t_y[u', f_u]$ such that for every $z \in S[u', f_u]$, $t_y[u', f_u](z) = D_T(F(y), u) - D_G(y, z)$. Since $t_x[w, f_u]$ and $t_y[u', f_u]$ must agree, it follows that there is a vertex $x' \in S[w, f_u]$, and a vertex $y' \in S[u', f_u]$ such that $t_x[w, f_u](x') + t_y[u', f_u](y') \geq D_G(x', y')$. By substituting for $t_x[w, f_u](x')$ and $t_y[u', f_u](y')$ we obtain

$$\begin{aligned}
D_T(F(x), F(y)) - D_G(x, y) &\geq (D_T(F(x), u) - D_G(x, x')) \\
&\quad + (D_T(F(y), u) - D_G(y, y')) - D_G(f(x'), f(y')) \\
&\geq 0.
\end{aligned}$$

Finally, suppose the statement of the claim holds for every $k' < k$ for some $k > 2$. We prove that the statement also must hold for $k$. We start by showing (2). Consider a vertex $x \in S_w$ such that for every $v_j$, $j \geq 1$ and every $y \in S_{v_j}$ we have that $D_T(F(x), v_j) - D_G(x, y) \leq 3d + 2$, that is, (2a) does not hold for $x$. We need to show that (2b) must hold for $x$. By the inductive hypothesis there is a type $t_x[v_3, f_{v_2}] \in \mathcal{L}[v_3, f_{v_2}]$ so that for every $y \in S[v_3, f_{v_2}]$, $t_x[v_3, f_{v_2}](y) = D_T(F(x), v_2) - D_G(x, y)$. Furthermore, by the assumption that (2a) does not hold for $x$, $t_x[v_3, f_{v_2}] \leq 3d + 2$. By the properties of succession of feasible $u$-states there must be a type $t'_x[v_2, f_u] \in \mathcal{L}[v_2, f_u]$ such that:

1. For every node $y$ in $(S[v_2, f_u] \cap S[v_3, f_{v_2}])$:

$$\begin{aligned}
t'_x[v_2, f_u](y) &= \beta(t_x[v_3, f_{v_2}](y) + 1) \\
&= \beta(D_T(F(x), v_2) - D_G(x, y) + 1) \\
&= \beta(D_T(F(x), u) - D_G(x, y)).
\end{aligned}$$

Observe that if $\beta(D_T(F(x), u) - D_G(x, y)) = \infty$ then $D_T(F(x), u) - D_G(x, y) > 3d + 2$ which implies that (2a) holds for $x$ which is a contradiction. Therefore, $\beta(D_T(F(x), u) - D_G(x, y)) \neq \infty$ so $\beta(D_T(F(x), u) - D_G(x, y)) = D_T(F(x), u) - D_G(x, y)$.

2. For every node $y$ in $S[v_2, f_u] \setminus S[v_3, f_{v_2}]$,

$$t'_x[v_2, f_u](y) = \beta \left( \max_{z \in S[v_3, f_{v_2}]} (t_x[v_3, f_{v_2}](z) + 1 - D_G(z, y)) \right).$$

As before, $t'_x[v_2, f_u](y) \leq 3d + 2$ because otherwise (2a) holds for $x$. Thus

$$t'_x[v_2, f_u](y) = \beta\left(\max_{z \in S[v_3, f_{v_2}]} (t_x[v_3, f_{v_2}](z) + 1 - D_G(z, y))\right)$$
$$= \max_{z \in S[v_3, f_{v_2}]} (t_x[v_3, f_{v_2}](z) + 1 - D_G(z, y)).$$

Following this

$$t'_x[v_2, f_u](y) = \max_{z \in S[v_3, f_{v_2}]} (D_T(F(x), v_2) - D_G(x, z) + 1 - D_G(z, y))$$
$$= D_T(F(x), u) - D_G(x, y).$$

Thus (2b) holds for $x$ and (2) is true for $|P| = k$. It remains to prove that (1) is true for $|P| = k$ as well. It is sufficient to prove that $F$ does not contract any $x \in (S_u \setminus S_{v_2})$ with any $y \in (S_w \setminus S_{v_{k-1}})$. There are two cases, either (2a) holds for $y$ or (2a) does not, and in that case, (2b) holds for $y$. In the latter case let $t_y[v_1, f_u]$ be the type in $\mathcal{L}[v_1, f_u]$ such that for every $z$ in $S[v_2, f_u]$, $t_y[v_1, f_u](z) = D_T(F(y), u) - D_G(y, z)$. Also, let $u' \in (N(u) \setminus v_1)$ be the neighbor of $u$ such that $x \in S[u', f_u]$. As in the proof of (1) for $k = 2$, let $t_x[u', f_u]$ be the type in $\mathcal{L}[u', f_u]$ such that for every $z$ in $S[u', f_u]$, $t_x[u', f_u](z) = D_T(F(x), u) - D_G(x, z)$. Again as in the proof of (1) for $k = 2$, $t_x[u', f_u](z)$ and $t_y[v_1, f_u]$ must agree which in turn implies that $F$ does not contract $x$ and $y$

To conclude, we consider the case when (2a) holds for $y$. Let $v_j$ be a vertex such that there is a $y' \in S_{v_j}$ so that $D_T(F(y), v_j) - D_G(y, y') > 3d + 2$. By the induction hypothesis, $F$ does not contract $x$ and $y'$. This gives us the following inequality for $D_T(F(x), F(y))$ and $D_G(x, y)$:

$$\begin{aligned} d_T(F(x), F(y)) &= D_T(F(x), v_j) + D_T(v_j, F(y)) \\ &\geq (D_T(F(x), F(y')) - D_T(v_j, F(y'))) + D_T(v_j, F(y)) \\ &\geq D_G(x, y') + D_G(y, y') + 3d + 2 - D_T(v_j, F(y')) \\ &\geq D_G(x, y) + 2d + 1 \geq D_G(x, y). \end{aligned}$$

This implies that the statement of the claim holds for every positive $k$. □

The claims together prove the existence of a non-contracting embedding $F$ of $G$ into $T$ with distortion at most $d$. □

**Proof of Theorem 1.3:** *Given a graph $G$, tree $T$ with maximum degree $\Delta$ and an integer $d$ we can decide whether $G$ can be embedded into $T$ with distortion at most $d$ in time $n^2 \cdot |V(T)| \cdot 2^{O((5d)^{\Delta^{d+1}} \cdot d)}$.*

*Proof.* The algorithm proceeds as follows. First, check that $\Delta(G) \leq \Delta^d$ (this follows from a local density argument). Now, we do bottom up dynamic programming on the tree $T$. For each vertex $u$ of the tree we make a Boolean table with an entry for each possible feasible $u$-state. For every leaf of the tree all the entries are set to true. For an inner node $u$ and a feasible $u$-state $\mathcal{X}_u$ we set $\mathcal{X}_u$'s entry to true if for each child $v$ of $u$ there is a feasible $v$-state $\mathcal{X}_v$ that succeeds $\mathcal{X}_u$ and so that $\mathcal{X}_v$'s entry is set to true. The algorithm returns "yes" if, at the termination of this procedure, there is a feasible $r(T)$-state $\mathcal{X}_{r(T)}$ with its table entry set to true. The algorithm clearly terminates, and correctness of this algorithm follows from Lemmas 4.10 and 4.9.

We now proceed to the running time analysis. In our bottom up sweep of $T$, we consider every edge and every vertex of $T$ exactly once, which yields a factor of $n_t = |V(T)|$. For each vertex $u$ we consider each feasible $u$-state $\mathcal{X}_u$ once, and for each such state and every child $v$ of $u$ of the state we need to enumerate all feasible $v$-states that succeed $\mathcal{X}_u$. In fact, we enumerate a larger set of candidate feasible $v$-states and for each such state $\mathcal{X}_v$ we check whether $\mathcal{X}_v$ succeeds $\mathcal{X}_u$.

First we show that the number of feasible $u$-partial embeddings is at most $n \cdot \Delta^{O(d^2 \cdot \Delta^{d+1})}$. This follows from the fact that for any vertex $u$ of the tree $|B(u, d+1)| \leq \Delta^{d+1}$ and that the domain of a any feasible $u$-partial embedding $f_u$ is contained in a ball of radius at most $2d+2$ in $G$. Because the degree of $G$ is bounded, a ball of radius $2d+2$ in $G$ can contain at most $\Delta^{O(d^2)}$ vertices.

One can easily prove that if the feasible partial embedding $f_u$ is given, the number of types and typelists that can appear in a feasible $u$-state together with $f_u$ is bounded by $(5d)^{\Delta^{d+1}}$ and $2^{O((5d)^{\Delta^{d+1}})}$ respectively. Thus, the number of feasible $u$-states is bounded by $2^{O((5d)^{\Delta^{d+1}} \cdot d)}$. If the domain $S_v$ of a feasible partial embedding $f_v$ for a child $v$ of $u$ is non-empty then we can use the fact that $S_v$ must have a non-empty intersection with the domain of $f_u$ to bound the number of potential successors of a $u$-state by $2^{O((5d)^{\Delta^{d+1}} \cdot d)} \cdot \Delta^d \leq 2^{O((5d)^{\Delta^{d+1}} \cdot d)}$. Since we can check whether a particular $u$-feasible state succeeds another in time $n \cdot 2^{O((5d)^{\Delta^{d+1}} \cdot d)}$ the overall running time of the algorithm is bounded by $n^2 n_t \cdot 2^{O((5d)^{\Delta^{d+1}} \cdot d)}$. $\square$